\documentclass[aps,prl,twocolumn,groupedaddress,floatfix]{revtex4}

\usepackage{graphicx,latexsym,bm}

\begin{document}

%\preprint{}

\title{Light Scattering in Transparent Glass Ceramics}


\author{S. Hendy}
\email{s.hendy@irl.cri.nz}
%\thanks{}
\affiliation{IRL Applied Mathematics, Industrial Research Ltd, Lower Hutt, 
New Zealand}


\date{\today}

\begin{abstract}
Transparent glass ceramic materials, with microstructures comprised of dispersed
nanocrystallites in a residual glass matrix, offer the prospect of nonlinear 
optical properties. However, good transparency requires low optical scattering and 
low atomic absorption. The attenuation of light due to scattering (turbidity) will 
depend upon the difference in refractive index of the two phases and the size and 
distribution of crystals in the glass. Here, we model the glass ceramic as 
a late-stage phase-separated structure, and compute scattering in this model. We find 
that the turbidity follows a $k^8 R^7$ relationship, where $k$ is the wavevector of 
light in the glass ceramic and $R$ is the average radius of the crystals in the glass.
\end{abstract}

%\keywords{}

\maketitle


Glass ceramics are glasses containing nanometer to micron sized crystals embedded in a glass matrix. Glass 
ceramics are easier to manufacture into complex shapes than traditional ceramics, utilizing standard techniques 
developed for the glass industry \cite{Beall92}. In addition, the embedded crystalline phase can enhance existing, 
or offer entirely new, properties from that of the parent glass. Recently, much attention has been paid 
to optically transparent glass ceramics \cite{Beall99,Tick95,Tick98}, which have greater thermal stability than 
their parent glasses. Transparent oxyflouride glass ceramics, for example, can be doped with optically active 
rare-earth cations, offering the possibility of numerous applications in nonlinear optics \cite{Tick95}. 
However, the understanding of the transparency of these nanophase glass ceramics is still relatively poor. 
Transparency has been found to occur in glasses with large volume fractions of crystals ($\sim$ 30-35 \%) and 
nanoscale crystal sizes (1-15 nm). Application of the Mie theory of scattering to such a material, leads to an 
over-estimate of the attenuation due to scattering (turbidity) by many orders of magnitude\cite{Beall99}.  

A more sophisticated approach is to use Rayleigh-Debye theory, which allows for the possibility of coherence and 
interference between scatterers. To describe the scattering, this theory requires a structure factor which depends 
on the distribution of scatterers in the medium. For example, Hopper \cite{Hopper85} has developed an 
approximate structure factor for glasses which have undergone late-stage microstructural phase separation via  
spinodal decomposition. Hopper's structure factor has been used to describe the scattering in transparent glass 
ceramics \cite{Tick95, Tick98} and gives a turbidity of  
\begin{equation}
\label{E1}
\tau \simeq 6.3 \times 10^{-4} ({\Delta n \over \bar{n}})^2 k^4 L^3
\end{equation}
where $k$ is the wavevector of the incident light propagating in a medium with refractive index $\bar{n}$,
$\Delta n$ is the difference in refractive index between the glass and crystal phases, and $L$ is the mean
distance between phases. This model improves on Mie theory, predicting turbidities that are somewhat closer 
to those measured for transparent glass ceramics. 

Subsequent to Hopper's work, there has been a considerable improvement in the understanding of 
late-stage structure factors in phase-separated media \cite{Langer,Furukawa85,Furukawa89a,Furukawa89b}. 
In particular, the small wave-number limit of the late-stage structure factor is now well characterized. We will 
apply this small wave-number structure factor to determine the attenuation due scattering in glass 
ceramics to improve on equation (\ref{E1}).

We consider the propagation of light in a medium with spatially varying dielectric constant $\epsilon(\bm{r})$.
We define $\eta(\bm{r})$ to be the fractional variation of the dielectric constant:
\begin{equation}
\label{E2}
\epsilon (\bm{r}) = \bar{\epsilon} \left( 1 + \eta (\bm{r}) \right)
\end{equation}
where $\bar{\epsilon}= \langle \epsilon \rangle$ is the average dielectric constant in the medium and 
$\langle \eta \rangle = 0$. We also define the two-point correlation function for the fractional variation in 
dielectric constant as follows:
\begin{equation}
\label{E3}
\phi (\bm{r}) = \langle \eta(\bm{r}+\bm{r}^\prime) \eta(\bm{r}^\prime) \rangle.
\end{equation}
The structure factor, $S(\bm{k})$, is just the Fourier transform of $\phi (\bm{r})$.

The Rayleigh-Debye approximation is valid in the scattering regime $k L \Delta \epsilon \ll 1$, where $\Delta \epsilon$ 
is the difference in dielectric constant between scatterer and background media \cite{kerker}. For optical wavelengths, 
with phase sizes of several tens of nanometers and $\Delta \epsilon \sim 0.3$, this quantity is typically less than 0.1. 
This approximation also neglects multiple scattering, which may become significant if the depth of the media becomes 
comparable to the inverse turbidity. 

In this approximation, the intensity of scattered light at an angle $\theta$ to unpolarized incident 
light, and the consequent turbidity, are given by:
\begin{eqnarray}
\label{E5}
{I(\theta) \over I_0} & = &  { k^4 \over 32 \pi^2 r^2} V \left( 1+\cos^2 \theta \right) S(2 k \sin \theta/2), \\
\label{E6}
\tau & = & {k^4 \over 16 \pi} \int^{\pi}_0 d \theta \sin \theta \left(1+\cos^2 \theta \right) S(2 k \sin \theta/2),
\end{eqnarray}
where $I_0$ is the intensity of the incident light, $V$ is the volume of the scattering media and $r$ is the distance 
from this media. 

To develop a structure factor for the variation in dielectric constant in glass ceramics, we will assume that 
the material is a binary mixture which has undergone late-stage phase separation. Thus, the local composition can 
be specified by a single parameter $\Phi = \varphi_A - \varphi_B$, where $\varphi_{A(B)}$ is the molar fraction of 
phase A (B). We also assume that the dielectric constant of the glass at any point is linearly related to the composition, 
$\epsilon = \epsilon_A \varphi_A + \epsilon_B \varphi_B$. The fractional deviation $\eta$ of the dielectric constant 
is then given by
\begin{equation}
\label{E7}
\eta (\bm{r}) = {\Delta \epsilon \over 2 \bar{\epsilon}} \left( \Phi (\bm{r}) - \langle \Phi \rangle \right)
\end{equation}
where $\langle \Phi \rangle$ is the average value of $\Phi$. 
Thus, the two-point correlation function (\ref{E3}), and hence the structure factor, is determined by the local 
composition, and will evolve as the composition evolves. 

The dynamics of such a system undergoing phase separation can be described by the Cahn-Hilliard equation 
\cite{cahn59}:
\begin{equation}
\label{E8} 
{\partial \Phi \over \partial t} = \bm{\nabla} \cdot \left( D \  \bm{\nabla} \mu (\Phi) \right), 
\end{equation} 
where $\mu (\Phi)$ is a generalized chemical potential for the two-phase system and $D$ is a diffusion coefficient. Some 
glass ceramics do phase separate via a spinodal mechanism (e.g. transparent mullite glass ceramics) while others 
nucleate homogeneously (e.g. transparent oxyflouride glass ceramics) \cite{Beall99}. While equation (\ref{E8}) 
cannot immediately be applied to nucleation\cite{Langer}, at sufficiently late times the structure of the phase
separated material will be independent of the initial separation mechanism\cite{Furukawa89a}. Hence, the 
late-stage structure of systems that have phase separated via nucleation can be approximated by solutions of 
equation (\ref{E8}).

\begin{figure}
\resizebox{8.5cm}{!}{\includegraphics{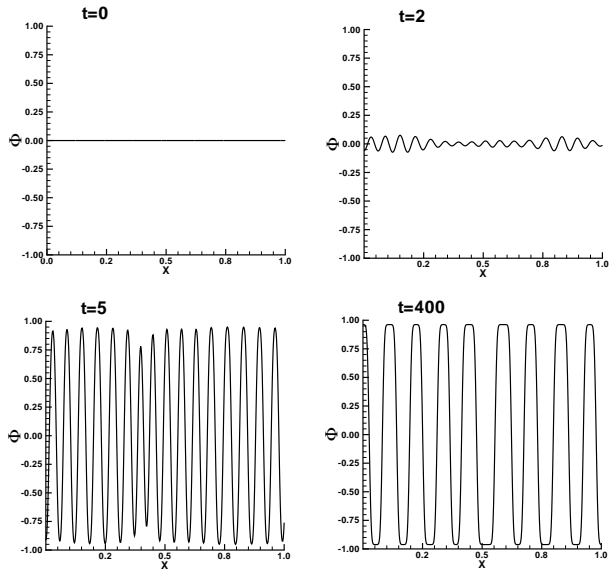}}
\caption{\label{spin4x4} The results of a 1D numerical simulation, using equation (\ref{E8}), of a system undergoing 
spinodal decomposition at early ($t=0,2,5$) and late times ($t=400$). At t=400, late-stage coarsening 
is evident.}
\end{figure}

Figure~\ref{spin4x4} shows a numerical solution of equation (\ref{E8}) in one spatial dimension for spinodal
decomposition. The early-stage spinodal instability produces a separation of phases from an initially 
homogeneous mixture. Late-stage coarsening leads to an increase in the size of the separated regions, as 
regions grow, evaporate and coalesce. This late-stage coarsening process, often called Ostwald ripening 
\cite{Ostwald}, occurs at late times in systems undergoing phase separation via both spinodal and nucleation 
mechanisms \cite{Voorhees92,Voorhees99}.

\begin{figure}
\resizebox{8.5cm}{!}{\includegraphics{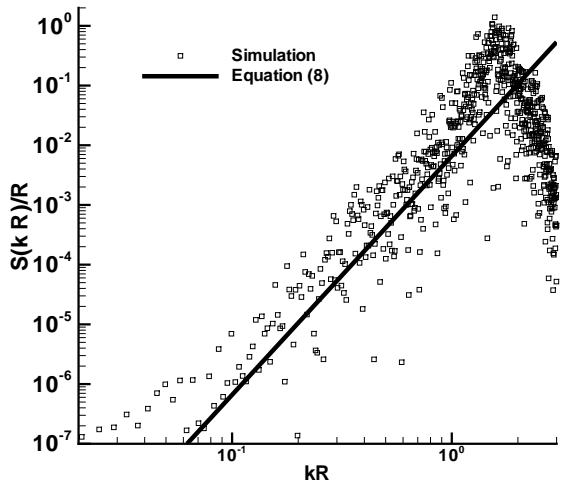}}
\caption{\label{structure} This figure shows the late-stage, small $kR$ structure factor for a single 1D 
numerical simulation of a system undergoing spinodal decomposition. The solid line shows the $(kR)^4$ power-law dependence 
for $kR < 1$, as given by (\ref{E9}) with $( \Delta \epsilon / 2 \bar{\epsilon})=1$ and 
$\varphi=0.2$. The noise is due to the random initial conditions, and finite spatial extent of the simulation. }
\end{figure}

The structure factor for a system undergoing late-stage coarsening via equation (\ref{E8}), can be approximated 
in the small $kR$ limit by:
\begin{equation}
\label{E9}
S(k) \simeq \varphi (1-\varphi) R^d \left( k R \right)^4 \left( {\Delta \epsilon \over 2 \bar{\epsilon}} \right)^2 
\end{equation}
where $R$ is the average crystal radius, $d$ is the number of spatial dimensions and $\varphi$ the crystal volume fraction 
\cite{Langer,Furukawa85,Furukawa89a,Furukawa89b}. This form of the structure factor has been observed in late-stage phase 
separated systems in computer simulations \cite{Furukawa89b} and experiment \cite{Gaulin}. Indeed, figure~\ref{structure} 
shows the late-stage structure factor for a 1D numerical simulation, where the $(kR)^4$ dependence is evident for $kR < 1$.

\begin{figure}
\resizebox{8.5cm}{!}{\includegraphics{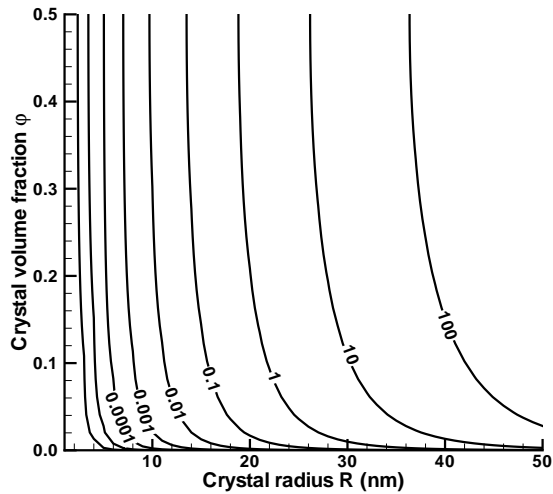}}
\caption{\label{turbidity} Contours of the turbidity given by equation (\ref{E11}) are shown as the 
crystal radius and crystal volume fraction are varied (with $\Delta n =0.1$, $\bar{n}=1.7$ and 
$\lambda_0 = 647$ nm).}
\end{figure}

\begin{figure}
\resizebox{8.5cm}{!}{\includegraphics{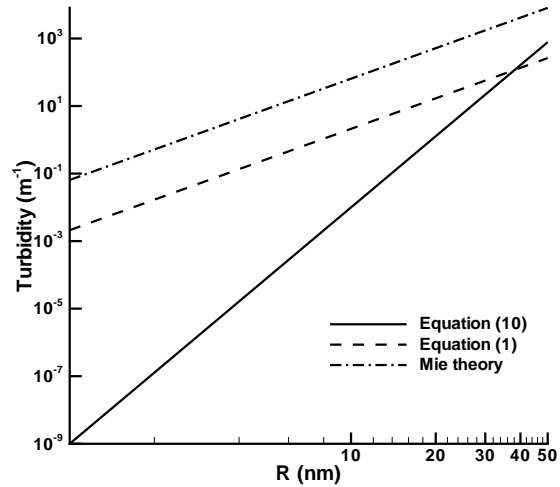}}
\caption{\label{tplot} The turbidities given by equation (\ref{E11}) (solid), equation
(\ref{E1}) (dashed) and by Mie theory (dash-dot) are compared as the crystal radius is varied (with 
$\varphi=0.3$, $\Delta n =0.1$, $\bar{n}=1.7$ and $\lambda_0 = 647$ nm).}
\end{figure}

Assuming that (\ref{E9}) holds, we can now give the scattered intensity and turbidity for a glass ceramic, with mean 
crystalline volume fraction $\varphi$, that has undergone late-stage coarsening:
\begin{eqnarray}
\label{E10}
{I(\theta) \over I_0} & \simeq & {V k^8 R^7 \over 2 \pi^2 r^2} \varphi (1- \varphi )  
\left( 1+\cos^2 \theta \right) \sin^4 \theta /2  \left( {\Delta n \over \bar{n}}\right)^2 \\ 
\label{E11}
\tau & \simeq &  {14 \over 15 \pi} \varphi (1-\varphi)  k^8 R^7 \left( {\Delta n \over \bar{n}} \right)^2
\end{eqnarray}
where we have assumed $\Delta \epsilon \ll \bar{\epsilon}$. 
In figure~\ref{turbidity}, contours of turbidity for (\ref{E11}) are shown as a function of crystal volume fraction and 
radius. In addition, figure~\ref{tplot} compares the turbidity predicted by Mie theory, equation (\ref{E1}) and equation 
(\ref{E11}), for a fixed volume fraction ($\varphi=0.3$). The strong $R^7$ dependence means that transparency is rapidly 
lost as R approaches $20-30$ nm. The $L^3$ dependence of Hopper's equation for turbidity (\ref{E1}) does not predict such 
a rapid transition. 

Mie theory is expected to be valid only if there are no spatial correlations in composition on a length-scale comparable 
to the wavelength of the incident light i.e. for early stage growth in the dilute limit ($L \gg R$). Equation (\ref{E11}) 
does not reduce to Mie theory in the dilute limit, as (\ref{E11}) assumes the presence of late-stage spatial correlations in 
composition. However, a theory of scattering that treats late-stage and early-stage structures via (\ref{E8}) will be sought 
in future work.  

Finally, we make comparison with turbidity data given by Tick, Borrelli, Cornelius and Newhouse \cite{Tick95}. They 
report on transparent oxyflouride glass ceramics with crystal radii $R \sim 15$ nm, crystal volume fractions $\phi=0.3$ 
and turbidities due to scattering of $2 \times 10^{-2}$ m$^{-1}$ for a wavelength of $\lambda_0=647$ nm (in vacuum). 
Assuming a mean refractive index $\bar{n}=1.7$ and a refractive index difference $\Delta n = 0.1$, Mie theory predicts 
a turbidity of $200$ m$^{-1}$, equation (\ref{E1}) predicts a turbidity of $7$ m$^{-1}$ (using $L \simeq 2.6 R$), 
whereas equation (\ref{E11}) predicts a turbidity of $0.2$ m$^{-1}$. Thus equation (\ref{E11}) represents a 
substantial improvement over (\ref{E1}) and Mie theory when applied to transparent glass ceramics.


\begin{thebibliography}{16}
\expandafter\ifx\csname natexlab\endcsname\relax\def\natexlab#1{#1}\fi
\expandafter\ifx\csname bibnamefont\endcsname\relax
  \def\bibnamefont#1{#1}\fi
\expandafter\ifx\csname bibfnamefont\endcsname\relax
  \def\bibfnamefont#1{#1}\fi
\expandafter\ifx\csname citenamefont\endcsname\relax
  \def\citenamefont#1{#1}\fi
\expandafter\ifx\csname url\endcsname\relax
  \def\url#1{\texttt{#1}}\fi
\expandafter\ifx\csname urlprefix\endcsname\relax\def\urlprefix{URL }\fi
\providecommand{\bibinfo}[2]{#2}
\providecommand{\eprint}[2][]{\url{#2}}

\bibitem[{\citenamefont{Beall}(1992)}]{Beall92}
\bibinfo{author}{\bibfnamefont{G.~H.} \bibnamefont{Beall}},
  \bibinfo{journal}{Annual Review of Materials Science}
  \textbf{\bibinfo{volume}{22}}, \bibinfo{pages}{91} (\bibinfo{year}{1992}).

\bibitem[{\citenamefont{Beall and Pinckney}(1999)}]{Beall99}
\bibinfo{author}{\bibfnamefont{G.~H.} \bibnamefont{Beall}} \bibnamefont{and}
  \bibinfo{author}{\bibfnamefont{L.~R.} \bibnamefont{Pinckney}},
  \bibinfo{journal}{Journal of the American Ceramics Society}
  \textbf{\bibinfo{volume}{82}}, \bibinfo{pages}{5} (\bibinfo{year}{1999}).

\bibitem[{\citenamefont{Tick et~al.}(1995)\citenamefont{Tick, Borrelli,
  Cornelius, and Newhouse}}]{Tick95}
\bibinfo{author}{\bibfnamefont{P.~A.} \bibnamefont{Tick}},
  \bibinfo{author}{\bibfnamefont{N.~F.} \bibnamefont{Borrelli}},
  \bibinfo{author}{\bibfnamefont{L.~K.} \bibnamefont{Cornelius}},
  \bibnamefont{and} \bibinfo{author}{\bibfnamefont{M.~A.}
  \bibnamefont{Newhouse}}, \bibinfo{journal}{Journal of Applied Physics}
  \textbf{\bibinfo{volume}{78}}, \bibinfo{pages}{6397} (\bibinfo{year}{1995}).

\bibitem[{\citenamefont{Tick}(1998)}]{Tick98}
\bibinfo{author}{\bibfnamefont{P.~A.} \bibnamefont{Tick}},
  \bibinfo{journal}{Optics Letters} \textbf{\bibinfo{volume}{23}},
  \bibinfo{pages}{1904} (\bibinfo{year}{1998}).

\bibitem[{\citenamefont{Hopper}(1985)}]{Hopper85}
\bibinfo{author}{\bibfnamefont{R.~W.} \bibnamefont{Hopper}},
  \bibinfo{journal}{Journal of Non-Crystalline Solids}
  \textbf{\bibinfo{volume}{70}}, \bibinfo{pages}{111} (\bibinfo{year}{1985}).

\bibitem[{\citenamefont{Langer}(1992)}]{Langer}
\bibinfo{author}{\bibfnamefont{J.~S.} \bibnamefont{Langer}}, in
  \emph{\bibinfo{booktitle}{Solids Far From Equilibrium}}
  (\bibinfo{publisher}{Cambridge University Press},
  \bibinfo{address}{Cambridge}, \bibinfo{year}{1992}),
  chap.~\bibinfo{chapter}{3}.

\bibitem[{\citenamefont{Furukawa}(1985)}]{Furukawa85}
\bibinfo{author}{\bibfnamefont{H.}~\bibnamefont{Furukawa}},
  \bibinfo{journal}{Advances in Physics} \textbf{\bibinfo{volume}{34}},
  \bibinfo{pages}{703} (\bibinfo{year}{1985}).

\bibitem[{\citenamefont{Furukawa}(1989{\natexlab{a}})}]{Furukawa89a}
\bibinfo{author}{\bibfnamefont{H.}~\bibnamefont{Furukawa}},
  \bibinfo{journal}{Journal of the Physical Society of Japan}
  \textbf{\bibinfo{volume}{58}}, \bibinfo{pages}{216}
  (\bibinfo{year}{1989}{\natexlab{a}}).

\bibitem[{\citenamefont{Furukawa}(1989{\natexlab{b}})}]{Furukawa89b}
\bibinfo{author}{\bibfnamefont{H.}~\bibnamefont{Furukawa}},
  \bibinfo{journal}{Physical Review B} \textbf{\bibinfo{volume}{40}},
  \bibinfo{pages}{2341} (\bibinfo{year}{1989}{\natexlab{b}}).

\bibitem[{\citenamefont{Kerker}(1969)}]{kerker}
\bibinfo{author}{\bibfnamefont{M.}~\bibnamefont{Kerker}},
  \emph{\bibinfo{title}{The Scattering of Light}} (\bibinfo{publisher}{Academic
  Press}, \bibinfo{address}{New York}, \bibinfo{year}{1969}).

\bibitem[{\citenamefont{Cahn and Hilliard}(1959)}]{cahn59}
\bibinfo{author}{\bibfnamefont{J.~W.} \bibnamefont{Cahn}} \bibnamefont{and}
  \bibinfo{author}{\bibfnamefont{J.~E.} \bibnamefont{Hilliard}},
  \bibinfo{journal}{Journal of Chemical Physics} \textbf{\bibinfo{volume}{31}},
  \bibinfo{pages}{688} (\bibinfo{year}{1959}).

\bibitem[{\citenamefont{Ostwald}(1900)}]{Ostwald}
\bibinfo{author}{\bibfnamefont{W.}~\bibnamefont{Ostwald}},
  \bibinfo{journal}{Zeitschrift f{\"{u}}r Physikalische Chemie}
  \textbf{\bibinfo{volume}{34}}, \bibinfo{pages}{495} (\bibinfo{year}{1900}).

\bibitem[{\citenamefont{Voorhees}(1992)}]{Voorhees92}
\bibinfo{author}{\bibfnamefont{P.~W.} \bibnamefont{Voorhees}},
  \bibinfo{journal}{Annual Review of Materials Science}
  \textbf{\bibinfo{volume}{22}}, \bibinfo{pages}{197} (\bibinfo{year}{1992}).

\bibitem[{\citenamefont{Alkemper et~al.}(1999)\citenamefont{Alkemper, Snyder,
  Akaiwa, and Voorhees}}]{Voorhees99}
\bibinfo{author}{\bibfnamefont{J.}~\bibnamefont{Alkemper}},
  \bibinfo{author}{\bibfnamefont{V.~A.} \bibnamefont{Snyder}},
  \bibinfo{author}{\bibfnamefont{N.}~\bibnamefont{Akaiwa}}, \bibnamefont{and}
  \bibinfo{author}{\bibfnamefont{P.~W.} \bibnamefont{Voorhees}},
  \bibinfo{journal}{Physical Review Letters} \textbf{\bibinfo{volume}{82}},
  \bibinfo{pages}{2725} (\bibinfo{year}{1999}).


\bibitem[{\citenamefont{Gaulin et~al.}(1987)\citenamefont{Gaulin, Spooner, and
  Morii}}]{Gaulin}
\bibinfo{author}{\bibfnamefont{B.~D.} \bibnamefont{Gaulin}},
  \bibinfo{author}{\bibfnamefont{S.}~\bibnamefont{Spooner}}, \bibnamefont{and}
  \bibinfo{author}{\bibfnamefont{Y.}~\bibnamefont{Morii}},
  \bibinfo{journal}{Physical Review Letters} \textbf{\bibinfo{volume}{59}},
  \bibinfo{pages}{668} (\bibinfo{year}{1987}).

\end{thebibliography}
\end{document}